\documentclass[paper]{ieice}
\usepackage[pdftex]{graphicx,xcolor}
\usepackage[fleqn]{amsmath}
\usepackage{newtxtext}
\usepackage[varg]{newtxmath}
\usepackage{url}

\usepackage[ruled]{algorithm2e}
\usepackage{appendix}

\usepackage{tikz}
\usepackage{etoolbox}
\usepackage{enumitem}
\newcommand*{\circled}[1]{\lower.7ex\hbox{\tikz\draw (0pt, 0pt)%
    circle (.5em) node {\makebox[1em][c]{\small #1}};}}
\robustify{\circled}

\field{}
\title{A Distributed Efficient Blockchain Oracle Scheme for Internet of Things}

\titlenote{The research was supported in part by the Guangxi Science and Technology Major Project (No. AA22068070), the National Natural Science Foundation of China (Nos. 62166004,U21A20474), the Key Lab of Education Blockchain and Intelligent Technology, the Center for Applied Mathematics of Guangxi, the Guangxi "Bagui Scholar" Teams for Innovation and Research Project, the Guangxi Talent Highland Project of Big Data Intelligence and Application, the Guangxi Collaborative Center of Multisource Information Integration and Intelligent Processing.}

\authorlist{%
\authorentry[xianyouquan@stu.gxnu.edu.cn]{Youquan Xian}{n}{1,2}\MembershipNumber{}
\authorentry[zhoulianghaojie@stu.gxnu.edu.cn]{Lianghaojie Zhou}{n}{1,2}\MembershipNumber{}
\authorentry[jiangyong@stu.gxnu.edu.cn]{Jianyong Jiang}{n}{1,2}\MembershipNumber{}
\authorentry[wangboyi@stu.gxnu.edu.cn]{Boyi Wang}{n}{1,2}\MembershipNumber{}
\authorentry[huohao@stu.gxnu.edu.cn]{Hao Huo}{n}{1,2}\MembershipNumber{}
\authorentry[liupeng@gxnu.edu.cn(Corresponding author)]{Peng Liu}{n}{1,2}\MembershipNumber{}
}
\affiliate[1]{The author is with the College of Computer Science and Information Technology, Guangxi Normal University, Guangxi, China.
}

\affiliate[2]{The author is with the Key Lab of Education Blockchain and Intelligent Technology, Ministry of Education, Guangxi Normal University, Guangxi, China.
}

\received{2015}{1}{1}
\revised{2015}{1}{1}

\begin{document}
\maketitle
\begin{summary}
In recent years, blockchain has been widely applied in the Internet of Things (IoT). Blockchain oracle, as a bridge for data communication between blockchain and off-chain, has also received significant attention. However, the numerous and heterogeneous devices in the IoT pose great challenges to the efficiency and security of data acquisition for oracles. We find that the matching relationship between data sources and oracle nodes greatly affects the efficiency and service quality of the entire oracle system. To address these issues, this paper proposes a distributed and efficient oracle solution tailored for the IoT, enabling fast acquisition of real-time off-chain data. Specifically, we first design a distributed oracle architecture that combines both Trusted Execution Environment (TEE) devices and ordinary devices to improve system scalability, considering the heterogeneity of IoT devices. Secondly, based on the trusted node information provided by TEE, we determine the matching relationship between nodes and data sources, assigning appropriate nodes for tasks to enhance system efficiency. Through simulation experiments, our proposed solution has been shown to effectively improve the efficiency and service quality of the system, reducing the average response time by approximately 9.92\% compared to conventional approaches.
\end{summary}


\section{Introduction}\label{sec1}

With the rapid development of the Internet of Things, the traditional centralized network structure faces many challenges, such as vulnerability to hacker attacks and paralysis caused by central failures. Blockchain, as a decentralized distributed network ledger technology, has advantages such as security, transparency, and traceability, which can well meet the needs of the Internet of Things\cite{wang2020blockchain}.

Blockchain can provide a secure and reliable underlying infrastructure for the Internet of Things. By using blockchain technology, the authenticity and uniqueness of IoT devices and data can be ensured, and data tampering can be prevented\cite{rahman2022blockchain}. However, blockchains and smart contracts have no access to the external systems (i.e., off-chain), nor use IoT devices to perform some computing tasks. Therefore, blockchain oracle is proposed to solve the above problems\cite{pasdar2023connect,ezzat2022blockchain,bodkhe2020blockchain}.

TEE is a security solution based on trusted hardware, which guarantees the security of the runtime environment by trusted hardware\cite{jauernig2020trusted}. Town Crier\cite{zhang2016town} is the first oracle scheme to combine Intel SGX with Ethereum smart contracts. Smart contracts can perform operations such as authentication and data acquisition processing in TEE, but Town Crier has a clear risk of single point of failure. Based on Intel SGX, Woo et al.\cite{woo2020distributed} proposed a decentralized oracle scheme for IoT applications. They are mainly used to collect data from IoT devices and improve the availability of oracle networks in the event of a single point of failure or a malicious node through the Byzantine Fault Tolerance (BFT) consensus mechanism. However, a key problem is that TEE devices are still relatively scarce in the Internet of Things, and a small number of devices are difficult to support the heavy computing and data acquisition tasks in the Internet of Things\cite{jauernig2020trusted}. In addition, we also analyze the aggregation strategy of threshold signature as an example. The conclusion shows that in a certain time range, the more nodes involved in obtaining data, the greater the probability of reaching a consensus. Therefore, it is necessary to improve the scalability of the oracle while ensuring security.

On the other hand, we find that the response time of the data obtained by the oracle is related to the choice of the oracle node, and there is also an implicit connection between the oracle node and the data source. For example, due to network and geographical differences, the response time of the German oracle node $A$ to obtain data from the German data source $D_1$ is usually faster than that from the Chinese data source $D_2$.
Although there are many oracle-related studies to improve the security and efficiency of the data obtained by the oracle. For example, based on the game theory to constrain the behavior of the oracle nodes\cite{peterson2015augur,adler2018astraea}, or through threshold signature and other cryptographic methods to ensure the credibility of the obtained data\cite{chainlink,dos}. However, how to improve the scalability while ensuring the security and credibility of the oracle system, and according to the inherent relationship between the node and the data source, select the appropriate node for the task to improve the efficiency of the oracle, is still an urgent problem to be solved. 

In this paper, we design a distributed and efficient blockchain oracle solution tailored for IoT scenarios, enabling the blockchain to acquire complex data or execute computational tasks from the IoT. Firstly, we design a novel distributed blockchain oracle architecture that incorporates hybrid TEE devices. By categorizing oracle nodes into ordinary nodes and TEE nodes, we ensure system security and trustworthiness while improving system scalability. Secondly, we define a new Quality of Service (QoS) standard for oracles to quantify the matching relationship between oracle nodes and data sources. We employ a weighted random algorithm to select more suitable nodes for each task, enhancing the execution efficiency of every task.

The contributions of this paper are summarized as follows :

   \begin{itemize}
        \item In order to improve the scalability of the oracle system while ensuring security and credibility, we designed a distributed multi-data source oracle architecture that includes TEE and ordinary devices. It uses a committee composed of TEE to improve the credibility and aggregation efficiency of off-chain data, and combines ordinary nodes to increase the scalability of the system.
        \item Since the selection of oracle nodes will affect the efficiency of data collection, we redesign a new quality of service (QoS) metric. It considers the response time and result accuracy of nodes obtaining data from the data source. Then, we use an efficient weighted random node selection algorithm to select task nodes in a non-replacement weighted extraction way, which fully improves the service efficiency of the system.
        \item Through experimental verification, our scheme can effectively improve the scalability and efficiency of the system. Compared with the ordinary scheme, the response time of each task is reduced by about 9.92\%.
    \end{itemize}

The rest of this paper is organized as follows. Section \ref{related_work} introduces some work related to this study. Section \ref{preparation} introduces oracle, threshold signature, and current problem. Section \ref{work} introduces our proposed oracle framework. Section \ref{result} shows the simulation results and analysis. Section \ref{conclusion} gives the conclusion of this paper.

\section{Related Work}
\label{related_work}
The purpose of the blockchain oracle is to help the blockchain obtain reliable off-chain data\cite{ezzat2022blockchain}. The data received by the oracle nodes may be inconsistent, and the oracle needs to ensure the authenticity and integrity of the acquired data. We divide the literature review into three parts: voting-based oracles,  hardware-based oracles, and distributed oracles.

\subsection{Voting-based Oracles}
Augur is the first prediction market platform for decentralized oracle proposed by Ethereum. It hands data request events to all participants in the market. If the data provided by one node is inconsistent with the consensus results of other nodes, the pledged tokens will be confiscated by the system and redistributed to other nodes, forcing the nodes to remain honest\cite{peterson2015augur}. Adler et al.\cite{adler2018astraea} designed a decentralised oracle scheme based on a voting game that divides participants into three categories: submitters, voters or verifiers. Voters and validators play games with each other. Eventually, the Nash equilibrium is reached where all rational participants are forced to act honestly. Cai et al.\cite{cai2020truth} thought that the linear growth of participants' equity may lead to herd behavior, so they propose a nonlinear equity scaling rule to prevent Sybil attacks. However, the voting-based oracle requires a definite answer to the proposition, and the user's voting is inefficient.

\subsection{Hardware-based Oracles}
Trusted Execution Environment (TEE) can protect the core code of the program from interference by other malicious programs, while avoiding data leakage\cite{jauernig2020trusted}. Zhang et al.\cite{zhang2016town} proposed a hardware-based oracle system, which uses Intel SGX to ensure the security and credibility of computing. However, the above system has the possibility of a single point of failure, so Woo et al.\cite{woo2020distributed,chen2021tora} proposed a distributed oracle using Intel SGX to improve the availability of the oracle network in the event of a single point of failure or malicious nodes through the BFT consensus mechanism. Liu et al.\cite{liu2022extending} using TEE to extend blockchain trust from the blockchain to the off-chain, continuously perceive the inside of the vaccine box through trusted sensors and generate anti-counterfeiting data to obtain reliable vaccine tracking data. However, hardware-based oracles require specific hardware support, so their scalability is poor and it is difficult to take advantage of the large number of IoT devices without TEE.

\subsection{Distributed Oracles}
DECO is a distributed oracle scheme that helps users prove that the data they access through TLS is from a specific website without the need for trusted hardware or server-side modifications\cite{zhang2020deco}. He et al.\cite{he2019sdfs} proposed a novel scalable data feed service SDFS, which ensures the rationality of the data acquisition process and the scalability of the system services. In addition, a node reputation evaluation strategy is designed to judge whether each node is a malicious node and update the reputation value of each node. Chainlink \cite{chainlink} proposes a decentralized oracle network to achieve reliable data input and connection between smart contracts and external data sources. It uses reputation systems and cryptographic algorithms to ensure reliable operation of the system. DOS Network\cite{dos} provides off-chain computing in a decentralized manner, and uses m-out-n multi-signature transactions to achieve consensus. Similarly, Manoj et al.\cite{manoj2023trusted} obtained trusted agricultural data for the blockchain using a threshold secret-sharing scheme. Lin et al.\cite{lin2022novel} proposed a novel IIoT-oriented digital twin architecture by combining oracle with decentralized learning. Gigli et al.\cite{gigli2023decentralized} proposed a robust oracle architecture with a reputation algorithm for selecting trusted data sources. Even if there are multiple malicious data sources, it can consistently provide high-quality data.

The above solutions are all focused on improving the accuracy, reliability, and completeness of oracle data. However, these solutions do not consider the impact of the matching relationship between oracle nodes and data sources on the performance of the oracle system. Therefore, this paper designs a high-performance distributed multi-data-source oracle system with a hybrid TEE device for IoT scenarios. By combining the advantages of TEE and distributed oracle systems, the system selects efficient and trustworthy oracle nodes through weighted random selection, thereby improving the access efficiency of the oracle system.

\section{Preliminaries and Current Problems}
\label{preparation}
\subsection{Oracle}
 The word `Oracle' comes from Greek mythology and refers to people who can directly communicate with God and foresee the future. In a blockchain environment, oracle is a system that provides blockchain with information from the real world\cite{caldarelli2020understanding}.
 
\begin{figure}[h!]
        \centering
        \includegraphics[width=3in]{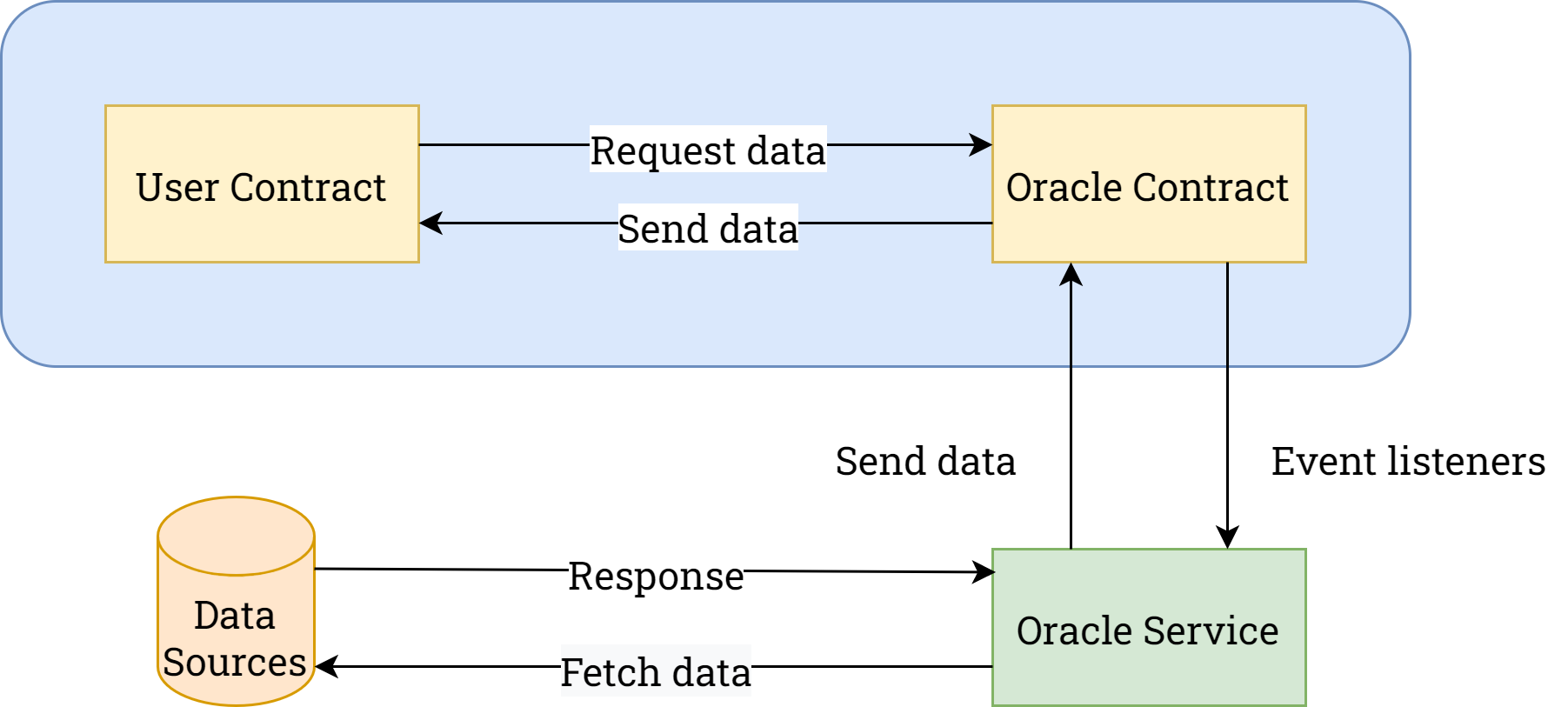}
        \caption{Overview of oracle system.}
        \label{fig:Oracle}
\end{figure}

 The main steps are shown in Fig. \ref{fig:Oracle}, 1) The user contract calls the oracle contract to obtain the data of the corresponding data source; 2) After the oracle node listens to the event of the oracle contract, it obtains data from the corresponding data source; 3) The data source returns the corresponding data; 4) Oracle nodes verify and sign the data and return the oracle contract; 5) The oracle contract returns data to the user contract.

 \subsection{Threshold Signature}
 Threshold signature is a group of $n$ members, the group has a pair of group signature public key and private key\cite{rong2015key}. The group's signature private key is shared by $n$ members in a threshold manner. Only the number of honest members in the group greater than or equal to the threshold value $t$ can sign on behalf of the group with the group signature private key. And it is easy for anyone to verify the signature through the group public key. These characteristics of threshold signature make it can be used in the process of oracle node aggregation data to prevent the problem of freeloading\cite{chainlink}. The process of threshold signature used in this paper is as follows:

 1) Generate public key and group member private key: input the security parameter $s$, the threshold value $t$ and the number of members $n$, and return the private key $sk_i (i=1,2,...,n)$ for each member's partial signature. Each member saves a private key fragment to generate a partial threshold signature. $pk$ is used to verify the threshold signature, as shown in the equation (\ref{eq:1}), where $L$ = { $sk_1,sk_2,...,sk_n$ }.
 \begin{equation}
    \label{eq:1}
        pk,L = Generate(s,t,n)
\end{equation}

2) Generate partial threshold signature: input the message msg and the member private key $sk_i$ to generate a partial signature $Sign_i$, as shown in the equation (\ref{eq:2}):
    \begin{equation}
    \label{eq:2}
        Sign_i = Sign(msg,sk_i)
    \end{equation}

3) Synthetic threshold signature: if the valid partial signatures $Sign_i$ collected by a member reach the threshold value $t$, these $Sign_i$ can be used to synthesize the threshold signature $Sign^*$.

4) Verify threshold signature: input message $msg$, public key $pk$ and threshold signature $Sign^*$, return judgment result $true$ and $false$, where $true$ means authentication passed and $false$ means not authenticated, as shown in equation (\ref{eq:3}):
    \begin{equation}
    \label{eq:3}
        Verify(msg,pk,Sign^*)==true
    \end{equation}

\subsection{Analysis of Existing Problems}

\subsubsection{Impact of number of nodes on aggregation success}
Facing the real-time data source in the Internet of Things, how to improve the probability of meeting the threshold $t$ in a certain time range is the core issue that we need to pay attention to. In this regard, we make a detailed analysis in Appendix \ref{model}. The conclusion shows that: 1) In order to get $m$ identical request results returned, the larger the number of selected nodes $n$, the higher the probability of reaching the threshold $m$; 2) The smaller the variation period $T$ of the data source, the more nodes $n$ are needed.
 
\subsubsection{The problem of matching data sources to oracles nodes}
\begin{figure}[h!]
    \centering
    \includegraphics[width=3in]{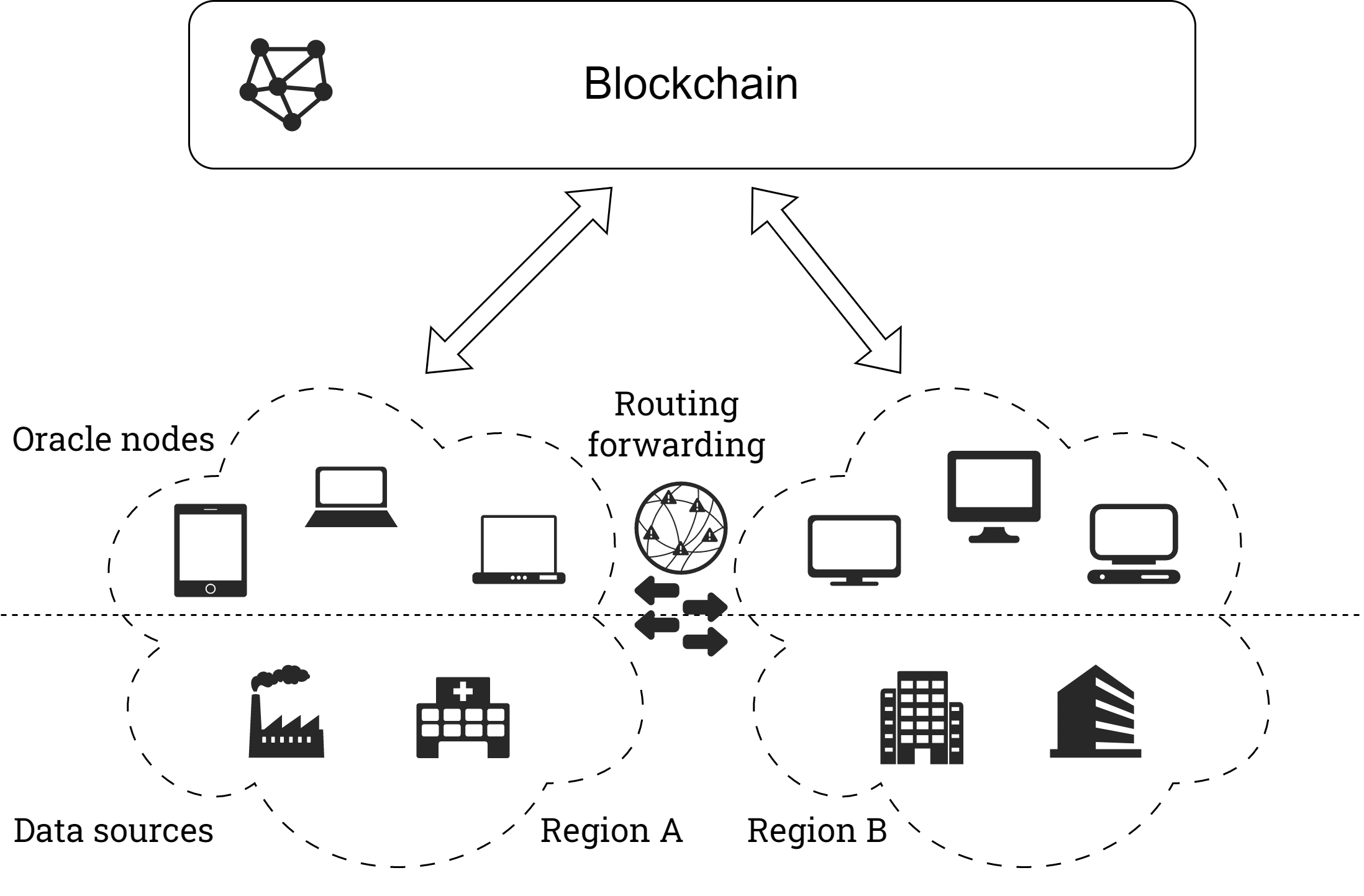}
    \caption{The process of obtaining data from data sources in regions A and B.}
    \label{fig:problem}
\end{figure}

In order to improve the robustness and security of the system, the current oracle scheme uses multiple oracle nodes and multiple data source nodes to reduce the impact of single point of failure and malicious\cite{gigli2023decentralized}. The user smart contract specifies multiple data sources to obtain data, and matches different oracle nodes to perform tasks. However, there is an implicit matching relationship between the oracle node and the data source. For example, the network connectivity will greatly affect the speed of data acquisition. As shown in Fig. \ref{fig:problem}, region A and B are distributed with some data sources and predictor nodes, and cross-domain requests will experience a longer network routing wait. Therefore, how to select appropriate nodes to participate in the task according to the implied matching relationship between data sources and oracle nodes is an effective means to improve the efficiency of the system.

\section{System Design}
\label{work}

\begin{figure*}[h]
    \centering
    \includegraphics[width=5in]{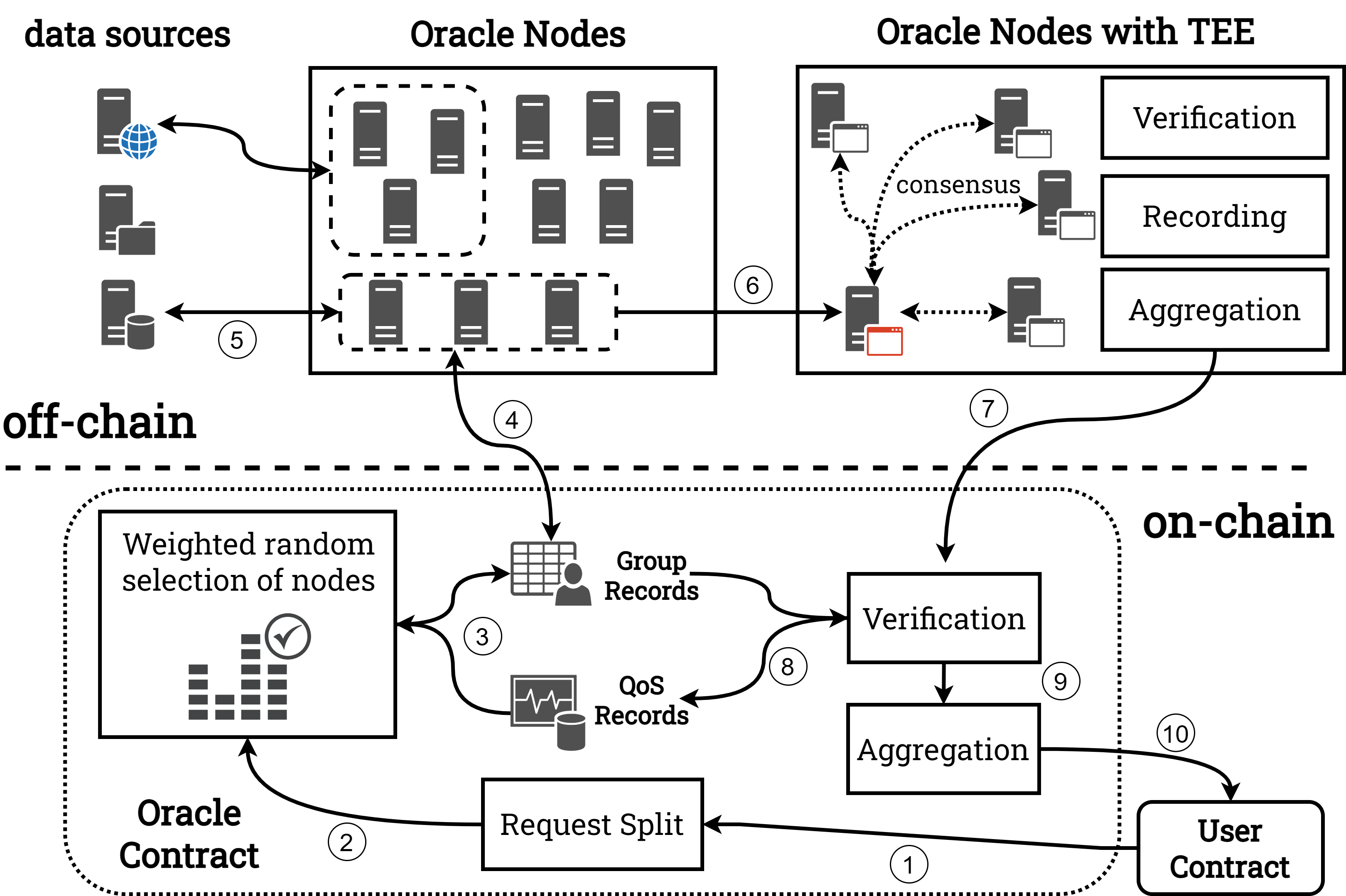}
    \caption{Overview of our oracle system.}
    \label{fig:Overview}
\end{figure*}

\subsection{Overview}
In this Section, we present the oracle scheme proposed in this paper. We design a high-performance distributed oracle weighted matching scheme with multiple data sources. The system is mainly divided into on-chain contracts and off-chain nodes, where off-chain nodes are further divided into ordinary oracle nodes and TEE nodes. Ordinary oracles account for the majority of industrial Internet devices, providing scalability for the oracle system. A few nodes with TEE can provide trusted support for the system and improve system trustworthiness. The system overview is shown in Fig. \ref {fig:Overview}.

The main steps include the following:
\begin{enumerate}[label=\circled{\arabic*}]
    \item Firstly, the user contract initiates a data acquisition request to the oracle contract.
    \item The oracle contract splits multiple data source requests into subtasks of a single data source.
    \item The oracle contract takes the QoS of each oracle node to the requested data source as the weight of node selection, and uses the A-EepJ  weighted random sampling algorithm to select the set of ordinary oracle nodes to participate in subsequent tasks.
    \item The off-chain oracle node uses a one-time non-interactive Distributed Key Generation (DKG) protocol to generate a distributed public and private key after listening to the task.
    \item According to the user task, the oracle node off-chain obtains data from the specified data source and signs it with its own private key fragment $sk_i$ to obtain the signature fragment $Sign_i$.
    \item The TEE node is responsible for jointly verifying the signature fragment $Sign_i$ uploaded by the ordinary node, recording its response time $t_d^i$, and finally returning the Leader, which is aggregated by the Leader to obtain the group signature $Sign^*$
    \item After aggregating the group signature, the Leader of the TEE node committee packages the group signature $Sign^*$, the original data and the node response time $t_d^i$, and then signs and uploads the oracle contract.
    \item After the signature of the oracle contract verification is correct, the original data is obtained, and the behavior information of the oracle node and the corresponding QoS record are updated.
    \item The final aggregation result is aggregated by oracle contract according to the user-defined multi-data source result aggregation strategy.
    \item Finally, the oracle contract returns the final result to the user contract.
\end{enumerate}

\subsection{Node Selection}

Firstly, the oracle contract processes data requests from user contracts in the form of a four-tuple $Request(m,n,D,r)$, where $m$ represents the number of nodes required for aggregate signature, $n$ represents the total number of nodes required for this task, $D$ represents the array of data sources, and $r$ represents the random number generated by the hash of the previous block. The oracle contract needs to split the multi-data source request task into multiple subtasks $Request(m,n,d,r)$ and record, $d \in D$, $d$ represents a single data source.
\begin{equation}
  Request(m,n,d,r) \gets Request(m,n,D,r)
\end{equation}
For a single subtask $Request(m,n,d,r)$, the oracle contract needs to select $n$ ordinary oracle nodes to obtain data d according to the requirements of the subtask. To select efficient nodes while ensuring randomness, we use the oracle node as a weight for the data source $d$ service quality $QoS_d$, and randomly select nodes. In order to reduce the space and time complexity of the oracle contract, we use the A-ExpJ algorithm \cite{efraimidis2006weighted} to randomly select $n$ nodes from $N$ to participate in the data acquisition task, $N$ represents the set of all ordinary oracle nodes. For each node $i$, a random number $u_i=rand(0,1)$ that obeys uniform distribution is selected, $k_i=u_i^{1/w_i}$ is used as the key value of sampling, and $m$ samples with the largest key value are selected as the sampling results. This method effectively reduces the time and space complexity of sampling compared with the cardinal method \cite{efraimidis2015weighted}. In addition, the non-replacement selection can reduce the repeated selection of some oracle nodes and reduce the response time.
\begin{equation}
    R = AExpJ(N,QoS_d,n)
\end{equation}

$R$ denotes the set of n ordinary oracle nodes selected from $N$ by the A-ExpJ algorithm. We give a detailed description in Algorithm \ref{algorithm:wrs}.

 \begin{algorithm}[h!]
    \caption{Weight node selection (AExpJ)}
    \label{algorithm:wrs}
    
    \LinesNumbered
    \KwIn {Set $V$, Weight of set elements $W$, Number of samples $m$}
    \KwOut {Selected Node Collection $R$}
    
    The first $m$ items of $V$ are inserted into $R$

    \For {$w_i \in R$}{
        $k_i = u_i^{1/w_i}$
        
        $u_i = random(0,1)$
    }
    
    Threshold $T_w$ is the minimum weight of $R$
    
    \While{$v_c \in V$ and $v_c$ is not been visited before}{
        $r = random(0,1)$
        
        $X_w = \frac{log(r)}{log(T_w)}$
        
        From the current item $v_c$ skip items until item $v_i$, such that:
        
        $w_c + w_{c+1} + ... + w_{i-1} < X_w \leq w_c + w_{c+1} + ... + w_{i-1} + w_i$
        
        The element in $R$ with the minimum weight is replaced by item $v_i$
        
       $t_w = T_w^{w_i}$
       
       $r_2 = random(t_w,1)$ 
       
       $k_i = (r_2)^{1/w_i}$
        
       Threshold $T_w$ is the minimum weight of $R$
    }
\end{algorithm}

\subsection{Data Request and Aggregation}

The oracle nodes are divided into two categories according to whether they have TEE: ordinary nodes and TEE nodes. TEE nodes together form a consensus committee to record, verify and aggregate data signatures. The ordinary oracle node generates a distributed public and private key based on the contract task, performs computational logic such as obtaining data or performing computational tasks, and signs the result.

Firstly, the selected oracle node set $R$ generates $n$ private key fragments $sk_i$ and public key $pk$ with a threshold of $m$ according to the $Request(m,n,d,r)$ one-time non-interactive distributed key generation (DKG) protocol  \cite{hanke2018dfinity} with a threshold of $m$.
 \begin{equation}
    pk,sk_i = Generate(r,m,n)
\end{equation}
Then, the node $i$ obtains data from the data source $d$, and uses its own private key fragment $sk_i$ to sign the obtained data $msg$ to obtain the signature fragment $Sign_i$.
\begin{equation}
    Sign_i = Sign(msg,sk_i)
\end{equation}

\begin{figure}[h!]
    \centering
    \includegraphics[width=3in]{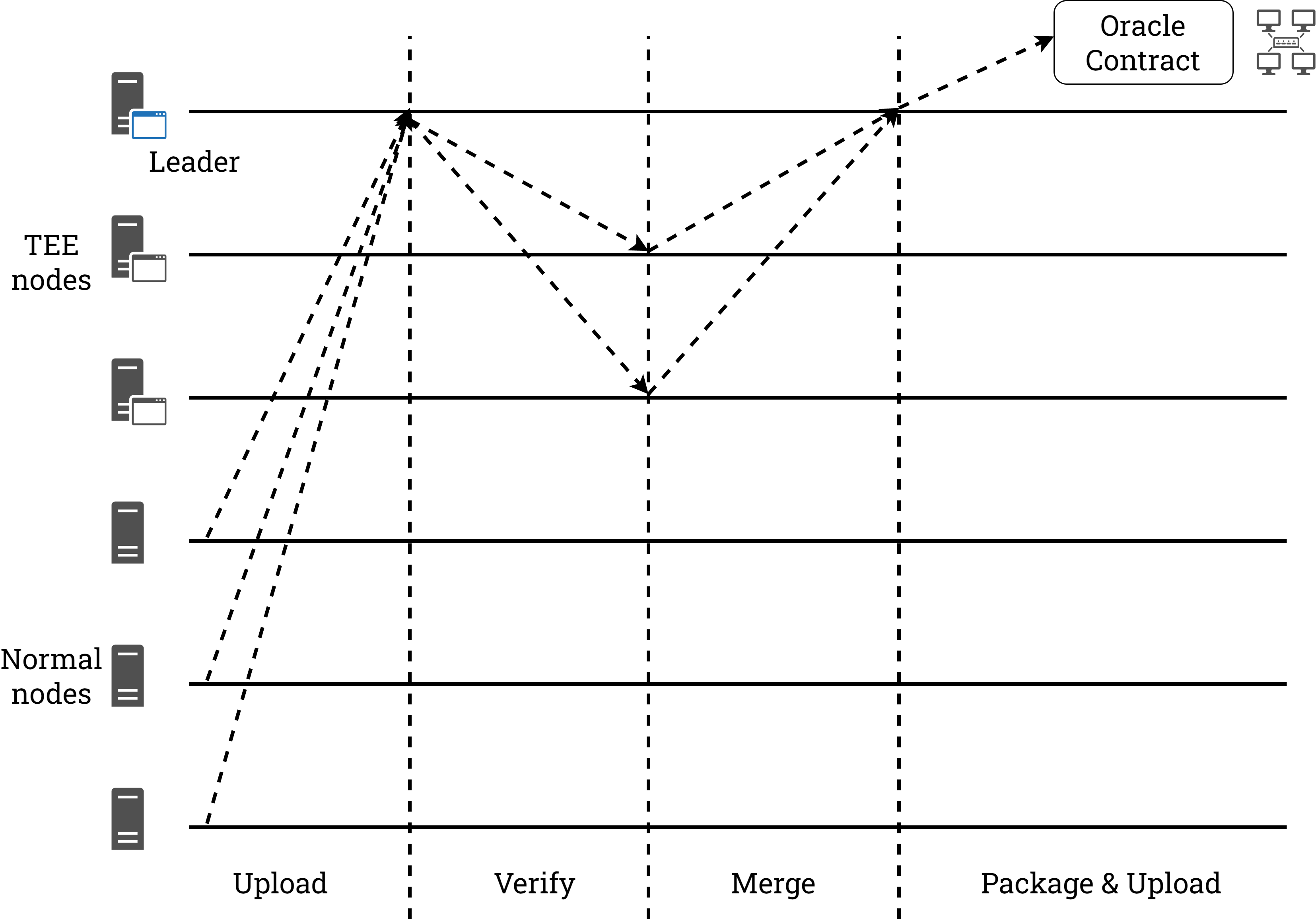}
    \caption{Data flow of TEE node committee.}
    \label{fig:TEE}
\end{figure}

As shown in Fig. \ref{fig:TEE}, the oracle node $i$ submits the signature fragment $Sign_i$ to the consensus committee formed by the TEE node. Their node executes the program in the TEE environment \cite{costan2016intel,hua2017vtz} to ensure the security and credibility of the program. The consensus committee verifies the signature fragment $Sign_i$ based on the Raft protocol \cite{ongaro2014search} and aggregates it to form a group signature $Sign^*$. In addition, the committee will also record the time when node $i$ uploads the signature as its response time $t^d_i$. Finally, the Leader node will package the consensus group signature $Sign^*$, the original data $msg$ and the successfully aggregated node response time $t^d$ as $S$, and use the Leader's private key $sk_{leader}$ to sign to get $sign_S$, upload the oracle contract.
\begin{equation}
    Sign_S = Sign(S,sk_{leader})
\end{equation}

The Raft-based TEE committee not only avoids the single point of failure issue of TEE nodes but also enhances resistance against Sybil attacks. Additionally, by using the Raft protocol for aggregated signatures, the number of broadcasts for threshold signature data is reduced, thus improving system efficiency. Furthermore, the TEE-based execution environment ensures the trustworthiness of response time records for regular nodes, providing a reliable data foundation for subsequent weighted random selection of nodes.

\subsection{Data Validation and QoS Update}

After receiving the data $Sign_S$ and $S$ from the TEE committee, the oracle contract uses Leader's $pk_{leader}$ to verify $Verify(S, Sign_S,pk_{leader})$, ensuring that the data has been verified by the TEE committee.

Then, the oracle contract uses the generated threshold signature public key $pk$ to verify the group signature $Sign^*$ formed by aggregation $Verify(msg,Sign^*,pk)$.

Finally, when the verification is passed, the oracle contract update the number of times $h$ and the response time $t_d^i$ that the node $i$ in $Sign^*$ successfully completes the task from the data source $d$.

In order to improve the quality of service of the oracle, we define a quality of service index of the oracle, which includes the response time and service accuracy of the oracle node $i$ to the request of the data source $d$. According to the current task results, we divide the device into nodes that complete the task and those that do not complete the task, and take the longest time $max(t_d)$ of the aggregated nodes as the total time $t_d^{total}$ of this task. At the same time, we use the timeout ratio $\beta$ of the device multiplied by the total time $max(t_d)$ of this task as the estimated time $t_d^{delay}$ of other unfinished tasks.
 \begin{equation}
    t_d^{delay} = \beta \times t_d^{total}
\end{equation}

We use the proportion of the response time $t_{d}^{i}$ of device $i$ to the longest time $t_d^{delay}$ of this task as its relative response time $T^d_{i} \in [0,1]$.
\begin{equation}
    T_d^{i} = 1 - \frac{t_{d}^{i}}{ t_d^{delay}}
\end{equation}

We use the number of times that the node $i$ successfully participates in the aggregation $h_d^i$ and the total number of times selected $c_d^i$ as the service accuracy $A_d^i$ of the node to the data source $d$.
 \begin{equation}
    A_d^i = \frac{h_d^i}{c_d^i}
\end{equation}

We set a hyper-parameter $\alpha$ to determine the proportion of response time and accuracy in the QoS of the node to the data source $d$. $\alpha > 0.5$ indicates that we pay more attention to the response time of the node, while $\alpha \leq 0.5$ indicates that we pay more attention to the service accuracy of the node.
\begin{equation}
    Qos_d^i = (\alpha * T_d^{i}) + ((1 - \alpha) * A_d^i)
\end{equation}

When we get the result returned by $Request(m,n,D,r)$'s $|D|$ subtask $Request(m,n,d,r)$, we try to aggregate it to get the final result $Result(s_1,...,s_{|D|})$ and return the user contract. It can be found that the user contract expects a consistent result for the $|D|$ results, but when the returned results are different in $s_1,...,s_{|D|}$ and it is difficult to reach an agreement, the user can set different aggregation strategies for multi-data source data according to their own needs.

\section{Experimental}
\label{result}

In this Section, we implemented a prototype of the system and conducted simulation experiments to validate the improvements in system efficiency and scalability offered by the proposed approach.

\subsection{Experiment setting}
The oracle system designed in this paper consists of on-chain contracts, off-chain oracle nodes, and TEE committee. The smart contract is written by Golang and deployed on the chainmaker blockchain\cite{chainmaker}, the on-chain and off-chain establish communication with the on-chain contract through the ChainMaker SDK. The total number of nodes $N$ is 10, the number of nodes $n$ selected for each task is 5, and the threshold $m$ required for each task is 3.

\subsection{Evaluation}
In order to validate the effectiveness of the proposed scheme, we will analyze how the proposed scheme improves the efficiency of the system and reduces the response time. Then, we also experimentally verify the impact of scalability on the success rate of data aggregation.

 \subsubsection{Response time}
\begin{figure}[h!]
    \centering
    \includegraphics[width=3in]{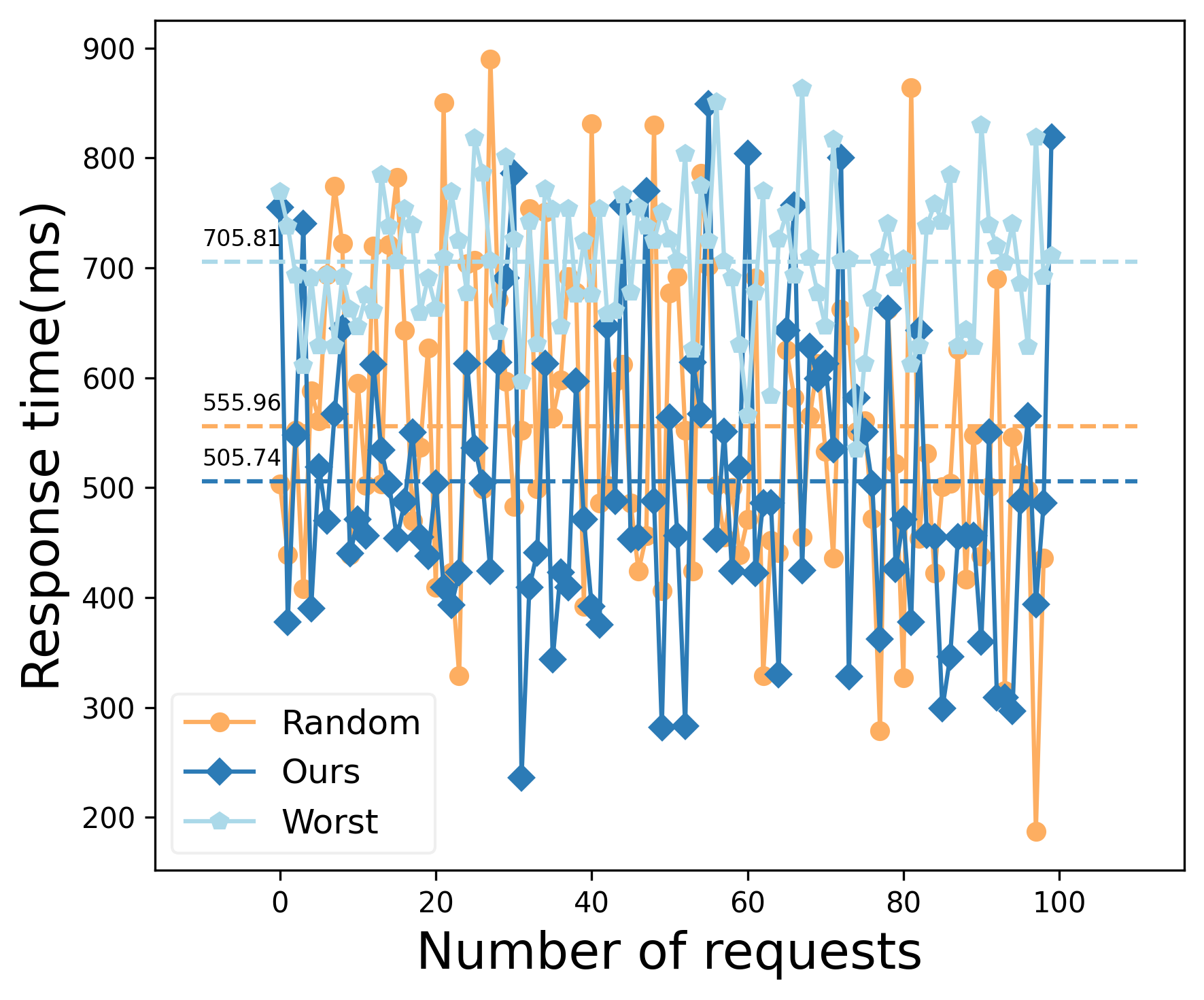}
    \caption{Response time of different strategies in 100 request tasks.}
    \label{fig:time}
\end{figure}

Fig.\ref{fig:time} shows the comparison of the response time of our scheme with the other two selection strategies (random selection and only the worst node) in 100 tasks. Obviously, the proposed scheme has a significant performance improvement compared with the traditional random node selection scheme, and the average response time per task is reduced by 9.92\%. Although the proposed scheme cannot achieve optimal performance, the weighted random method can maintain a certain random selection to prevent specific attacks against high-reputation users. This is consistent with our original intention to improve the performance of the oracle system while maintaining security.

\subsubsection{Effectiveness of QoS}
      \begin{figure}[h!]
    \centering
    \includegraphics[width=3in]{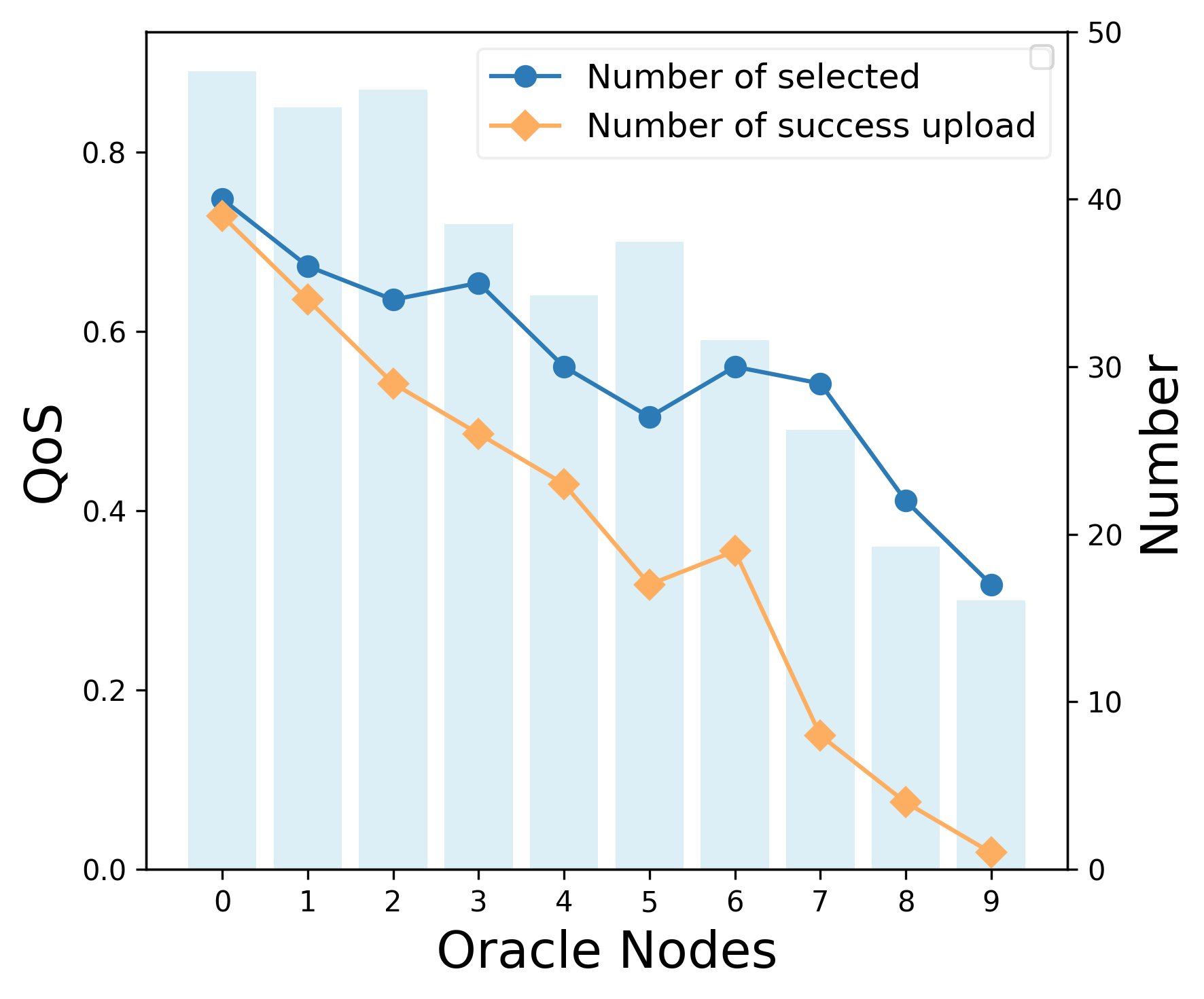}
    \caption{Effectiveness analysis of QoS computing.}
    \label{fig:QoS}
\end{figure}
Fig.\ref{fig:QoS} shows the change of metrics after our scheme performs 100 request tasks on a data source on 10 prophetic nodes. Obviously, it can be observed that there is a clear correlation between the node QoS and the successful completion of the task. This shows that the proposed scheme can identify nodes with slow response and reduce their QoS according to the actual operation status of nodes. This is consistent with our expectations. The higher the QoS of the high-performance nodes that can complete the task in time, the greater the probability of their selection.

\subsubsection{Node selection and QoS}
\begin{figure}[h!]
    \centering
    \includegraphics[width=3in]{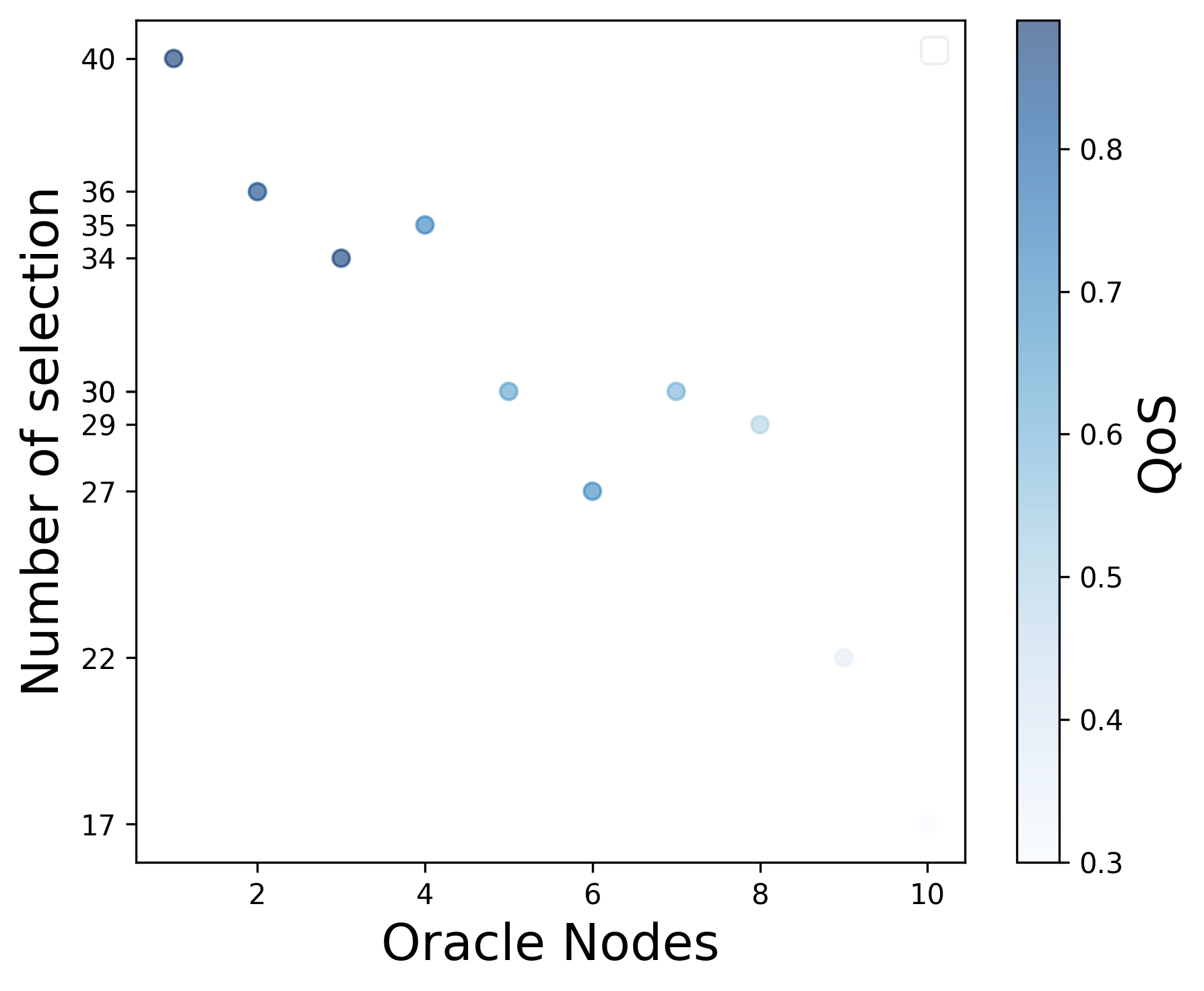}
    \caption{The Relationship between node selection and QoS.}
    \label{fig:node_seletced}
\end{figure}

In addition, we also analyze the relationship between QoS and the number of nodes selected more intuitively through the Fig. \ref {fig:node_seletced}. We can find that with the decrease of QoS, the number of node selection is also decreasing. This shows that the proposed node selection scheme is effective. 

 \subsubsection{Expandability}
\begin{figure}[h!]
    \centering
    \includegraphics[width=3in]{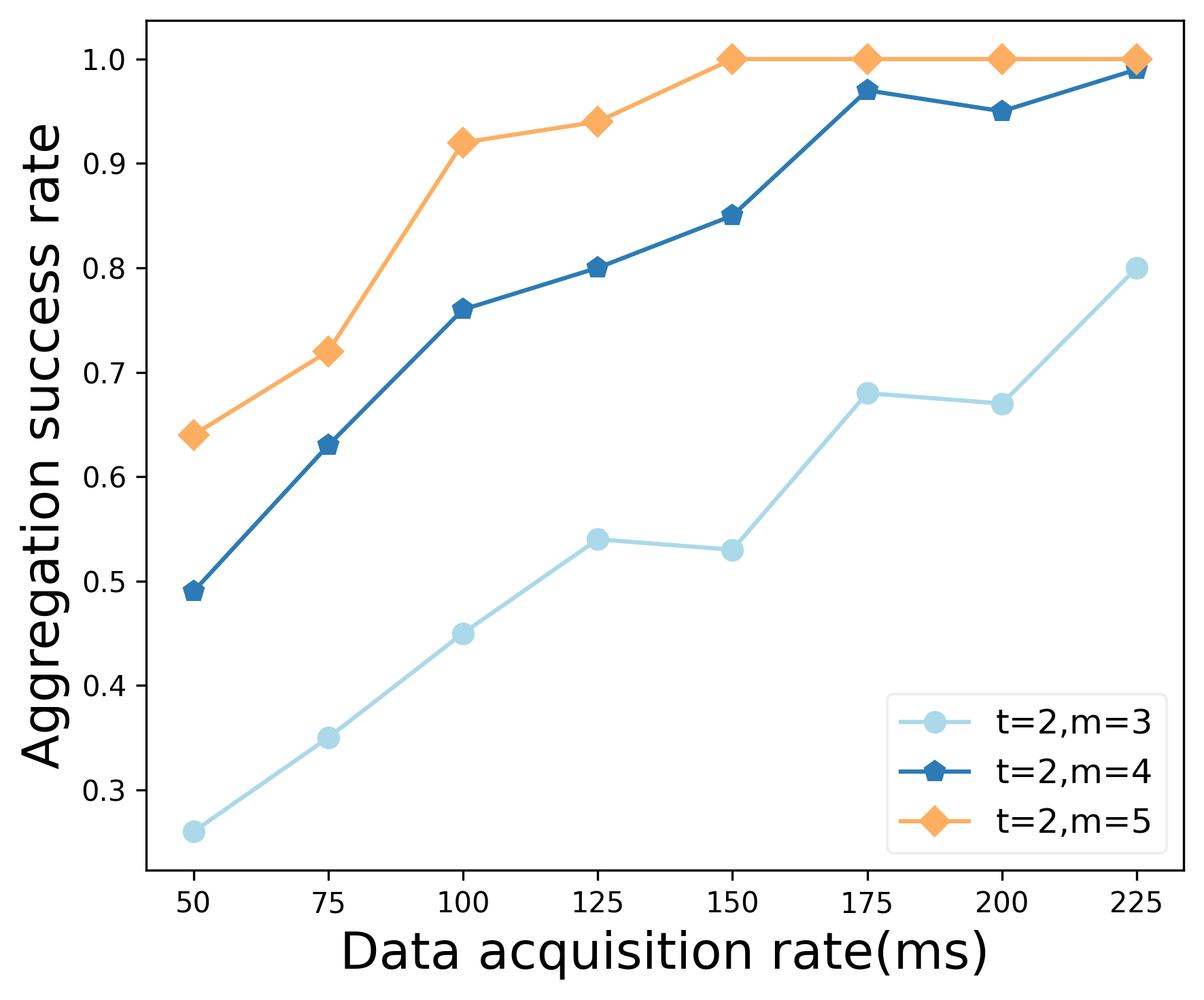}
    \caption{The relationship between aggregation success rate and the number of participating nodes.}
    \label{fig:data_acquisition}
\end{figure}

Fig. \ref{fig:data_acquisition} shows the impact of system scalability on the success rate of data aggregation in real-time data acquisition tasks in the Internet of Things. We observe the effect of choosing a different number of nodes $m$ per round on the success rate of oracle aggregation with a threshold $n$ of 2. The result shows that selecting more nodes per round of tasks is more likely to result in successful aggregation under the same threshold. This is consistent with the original intention of our design, and it is also the reason why we design the hybrid TEE oracle architecture. Massive ordinary nodes are responsible for data acquisition tasks to improve system performance. A small number of secure TEE nodes form a fault-tolerant recording and verification committee, which is responsible for ensuring the security and credibility of data.

\section{Conclusions}
\label{conclusion}
This paper proposes a distributed efficient blockchain oracle scheme for the Internet of Things, which aims to obtain IoT data efficiently and credibly for the blockchain. In order to improve the success rate of aggregation data when the oracle obtains real-time data, we divide the nodes into two categories, and combine the advantages of ordinary nodes and TEE nodes to improve the scalability and aggregation success rate. Then we propose a new oracle quality of service standard to measure the relationship between nodes and data sources, select more suitable nodes for tasks, and improve system performance. Experiments show that the scheme can effectively improve the access efficiency and scalability of the system, so that the blockchain can obtain IoT data more efficiently. 

\appendix
\section{Modeling of the problem}
\label{model}
    \subsection{Hypothesis}
\begin{figure}[h!]
        \centering
        \includegraphics[width=3in]{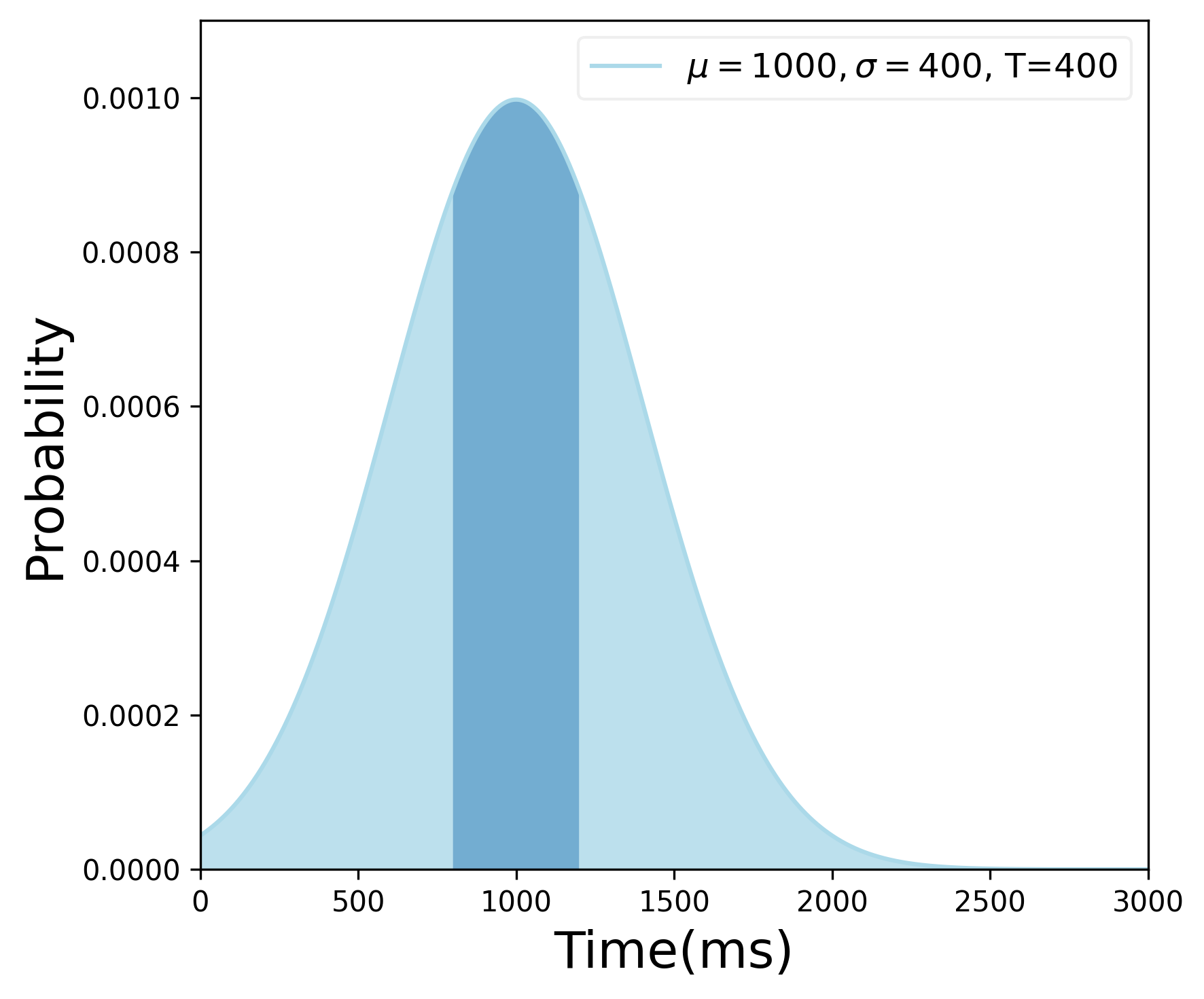}
        \caption{The Relationship between Node Selection and QoS.}
        \label{fig:distribution}
    \end{figure}

    We assume that $n$ selected oracle nodes are given the data request task at the same time, their response times satisfy the $(\mu,\sigma)$ normal distribution, and the collection of data from IoT devices satisfies a periodic variation with a period of $T$\cite{fraleigh2003provisioning}. As shown in the Fig. \ref{fig:distribution}.

    \subsection{Question}
     The problem is defined as: in the normal distribution curve in the interval of $[0,+\infty]$ and x axis surrounded by the area of $S_{total}$, Randomly scattered $n$ points, the probability that the number of scattered points $m$ in the $[x,x+T]$ interval area $S_{t}$ is greater than or equal to the threshold value $t$ of the threshold signature.

    \subsection{The relationship between threshold requirement and $n$}
      The total probability interval area $S_{total}$ is expressed as:
          \begin{equation}
        S_{total} = \int_{0}^{+\infty}N(\mu,\sigma^2)dx
    \end{equation}
    
    Because of $\mu \ge 0$, so $\frac{1}{2} \leq S_{total} \leq 1$.
    
    The selected probability interval $S_{t}$ in $T$ time is:
    \begin{equation}
        S_{t} = \int_{t}^{t+T}N(\mu,\sigma^2)dx
    \end{equation}
    Therefore, when test or select of 1 node, the probability of occurrence of event $A$ is $P$ $(0<p<1)$:
      \begin{gather}
         P = \frac{S_{t}}{S_{total}}
    \end{gather}
    In $n$ independent experiments, the probability that event $A$ happens to occur $m$ times is:
    \begin{equation}
        P_n(m) = C_n^m P^m(1-P)^{n-m}
    \end{equation}
    We transform the problem into an N-fold Bernoulli experiment, conduct $n$ experiments or select $n$ nodes, and the probability of successfully aggregating group signatures varies with $n$ as follows($P_n(m \ge t)$ is the maximum probability that the threshold $t$ can be met):
    \begin{equation}
        P_n(m \ge t) = \sum_{i=t}^{n} C_n^i P^i(1-P)^{n-i}
    \end{equation}
    Since both $S_t$ and $S_{total}$ are independent of $n$, $P$ is also independent of $n$.
    
    Also, $C_n^i = \binom{n}{i} = \frac{n!}{i!(n-i)!} $is a positive correlation with $n$.
    
    And $P_n(m\geq t)$ is an additive term. When $n$ increases, $P_n(m\geq t)$ increases monotonically.
    
    Therefore, $n$ is positively correlated with $P$, and when  $n$ increases, $P$ also increases.

    \subsection{Conclusion}
    Therefore, we can get a conclusion that when the number of nodes $n$ selected for each task increases, the probability $P$ of successful data aggregation obtained in changing data such as the Internet of Things also increases.

\bibliographystyle{ieicetr}
\bibliography{myref}

\profile[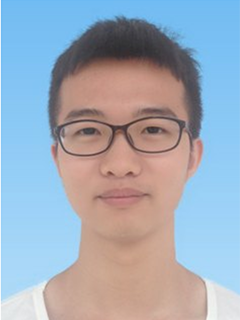]{Youquan Xian}{received the BEdegree from the BeiBu Gulf University, in 2021. He is currently working toward the master’s degree at Guangxi Normal University. His research interests include blockchain, edge computing, and federated learning.}
\label{profile}

\profile[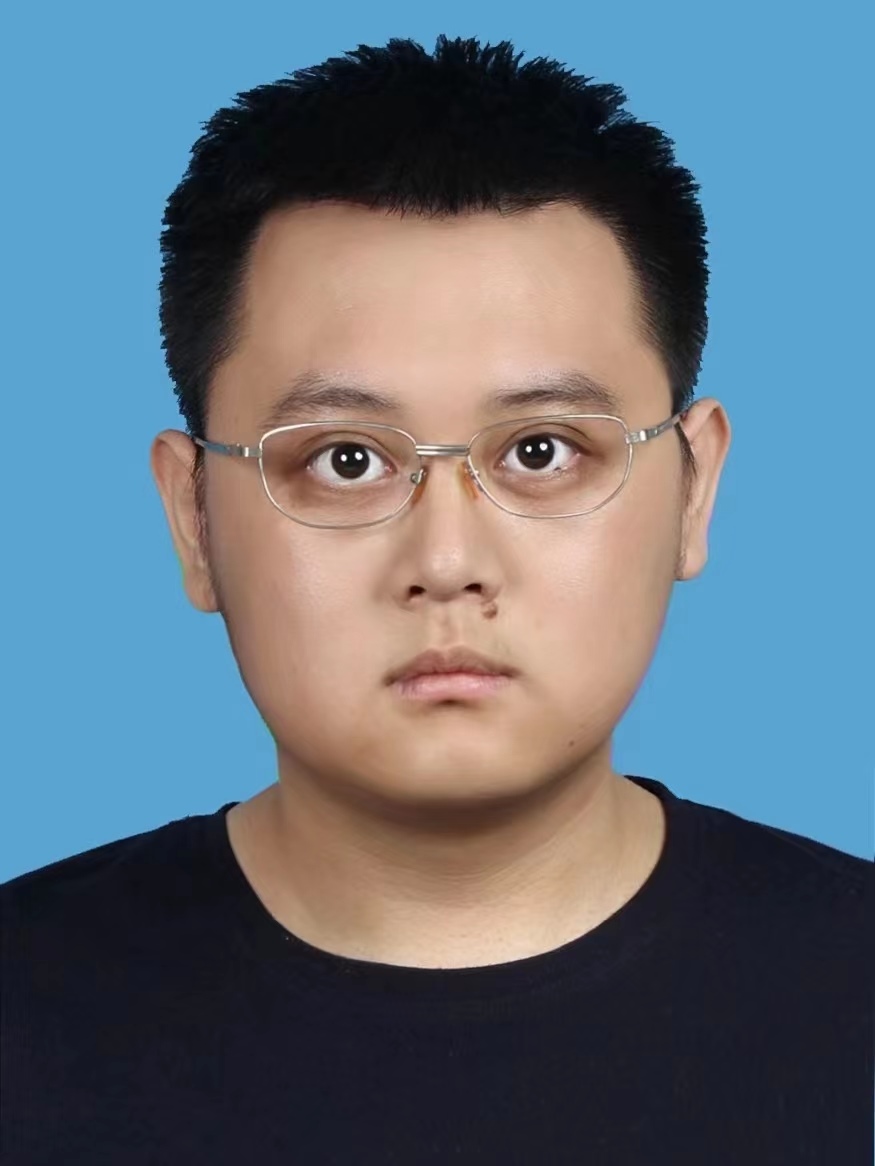]{Lianghaojie Zhou}{received the BEdegree from the HuaQiao University, in 2019. He is currently working toward the master’s degree at Guangxi Normal University. His research interests include blockchain and energy trading.}

\profile[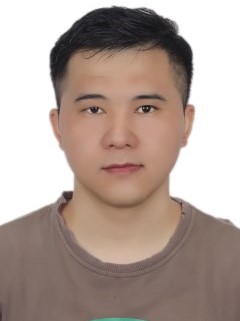]{Jianyong Jiang}{is currently pursuing a master's degree at Guangxi Normal University. His research encompasses various aspects, including distributed ledger technology, consensus algorithms, and smart contracts.}
\label{profile}

\profile[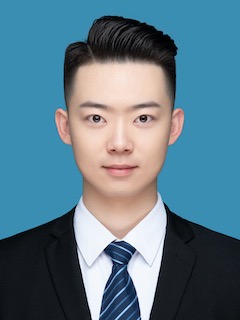]{Boyi Wang}{received the BEdegree from the Chongqing Jiaotong University, in 2020. He is currently working toward the master’s degree at Guangxi Normal University. His main interests are blockchain and privacy computing.}
\label{profile}

\profile[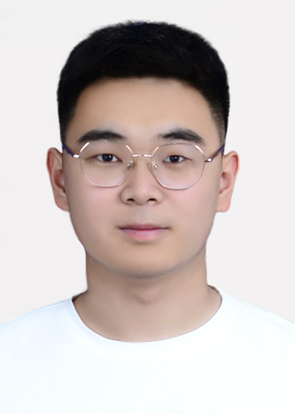]{Hao Huo}{obtained a bachelor 's degree in 2021 from Shanxi Agricultural University. He is currently studying for a master 's degree from Guangxi Normal University. His research interests include blockchain and cryptography, and he is committed to the privacy protection of blockchain.}
\label{profile}

\profile[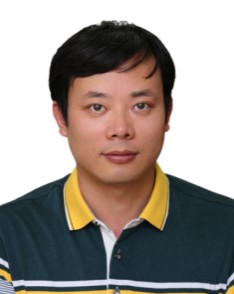]{Peng Liu}{received his Ph.D. degree in 2017 from Beihang University, China. He began his academic career as an assistant professor at Guangxi Normal University in 2007 and was promoted to full professor in 2022. His current research interests are focused on federated learning and blockchain.}
\label{profile}

\end{document}